\documentclass[12pt, a4paper]{article} 
\usepackage{times} 
\usepackage{amsmath} 
\usepackage{latexsym} 
\usepackage{mathrsfs} 
\begin{document} 
\vglue1cm 
\thispagestyle{empty} 
\begin{center}{\Large{\bf On f(R) theories in two-dimensional spacetime}}\end{center} 
\vglue 0.5cm 
\begin{center} {\large {\bf M. A. Ahmed$^a$}}\end{center} 
 
\begin{center} {\large{Physics Department, Kuwait University, Kuwait}}\end{center} 
\vglue1cm 
\hglue-0.65cm{\bf Abstract}. In recent years, theories in which the 
Einstien-Hilbert lagranigan is replaced by a function $f(R)$ of the Ricci 
Scalar have been extensively studied in four-dimensional spacetime.  In this 
work we carry out an analysis of such theories in two-dimensional spacetime 
with focus on cosmological implications.  Solutions to the cosmological field 
equations are obtained and  their properties are analysed.  Inflationary 
solutions are also obtained and discussed. Quantization is then carried out, 
the Wheeler-DeWitt equation is set up and its exact solutions obtained.\\

\vglue3cm 
\hglue-0.63cm 
${^a}email: mahmed@kuc01.kuniv.edu.kw$ 
\newpage 
\setcounter{page}{1} 
\section {Introduction} 
Attempts to modify the theory of general relativity, by including higher order 
invariants in the action, started not too long after its inception [1], [2]. 
Later the non-renormalizability of general relativity gave impetus to the 
inclusion of higher order terms in the action [3].  More recently it was shown 
that when quantum corrections are taken into consideration, higher order 
curvature invariants need to be added to the low energy gravitational action 
[4].  Such considerations further increased the interest in constructing 
theories in which the Einstien-Hilbert action is extended by the inclusion of 
higher order curvature invariants with respect to the Ricci Scalar.  Our 
interest here is in the so-called $f(R)$ theories of gravity.  In these 
theories the Lagrangian in the Einstein-Hilbert action  
 
\begin{equation}\label{eq:1} 
I'_G = -\frac{1}{2\kappa} \int { d}^4 { x} \sqrt{-g}~ R~, 
\end{equation} 
  where $\kappa$ = 8$\pi G$, $G$ is the gravitational constant, $g$ is the determinant of the metric tensor and $R$ is the Ricci scalar (in units $c$ = $\hbar$ = 1), is generalized to become 
 
\begin{equation}\label{eq:2} 
I' = -\frac{1}{2 \kappa} \int {d}^4{x} \sqrt{-g}~ f(R)~. 
\end{equation} 
In eq.\eqref{eq:2}  $ f(R)$ is a general function of $R$ [5].  Our focus here is on the cosmological aspects of $f(R)$ theories.\\ 
Now in another direction, the quest for quantum theory of gravity has led to 
the study of the simpler case of gravitational theory in two-dimensional 
spacetime.  Such a spacetime provides an interesting arena in which to explore 
some fundamental aspects of both classical and quantum gravity. The reduction 
in the degrees of freedom greatly simplifies the analysis of the field 
equations. This leads to appreciable understanding of several problems in 
gravity theory. In two-dimensional spacetime,  the two-dimensional gravitational constant $G_2$ is dimensionless and formally the theory with the bare action is 
 
\begin{equation}\label{eq:3} 
I_G = -\frac{1}{2 g_{_N}} \int{ d}^2 {x}\sqrt{-g}~ R~, 
\end{equation} 
where $g_{_N}$  = 8$\pi G_2$, is power counting renormalizable in perturbation theory.  However the Einstein$-$ Hilbert action term is purely topological in two dimensions.  In fact in two spactime dimensions, the curvature tensor R$_{\mu\nu\lambda\rho}$ has only one independent component since all nonzero components may be obtained by symmetry from {\textit R$_{0101}$.  Equivalently the curvature tensor may be written in terms of the curvature scalar [6]

\begin{equation}\label{eq:4} 
R_{\mu\nu\lambda\rho} =\frac{1}{2} R~ ( g_{\mu\lambda} g_{\nu\rho} - g_{\mu\rho} 
g_{\nu\lambda})~, 
\end{equation} 
so that $R$ alone completely characterizes the local geometry.  Eq.\eqref{eq:4} implies that 
\begin{equation}\label{eq:5} 
R_{\mu\nu} =\frac{1}{2} g_{\mu_\nu} R~, 
\end{equation} 
so that the Einstein tensor {\it G}$_{\mu\nu}$  = {\it R}$_{\mu\nu}$ - $\frac{1}{2}{\it g}_{\mu\nu} {\it R}$,  vanishes identically and the usual Einstein equations are meaningless in two dimensions.  This led to various models for gravity in two-dimensional spacetime being proposed [7].  Of special interest are those models that involve a scalar field, the dilation, in the action [7] $-$ [10].  We have previously studied some aspects of classical and quantum cosmology in two-dimensional dilation gravity models [11], [12].  In the present work we study {\it f(R)} theories as an alternative way to formulate gravitational theory in two-dimensional spactime and explore some of their cosmological implications.  We restrict ourselves to the classical theory only. \\ 
In sect. 2 we set up the {\it f(R)} gravity theory in two-dimensional 
spacetime and derive the general field equations.  We then specialize to the 
case of the Friedmann $-$ Robertson $-$Walker metric and obtain the field 
equations with matter treated as a perfect fluid.  Sect. 3 is devoted to 
obtaining solutions to the cosmological field equations under various 
conditions of matter or radiation dominance.  Properties of these solutions 
are discussed in sect. 4.   In particular, conditions for ensuring cosmic 
acceleration and solving the horizon problem are elucidated.  Inflation is 
discussed in sect. 5 and solutions to the field equations in the absence of 
matter or radiation are obtained and their properties are discussed. In sect.6 
we carry out the quantization. We establish the Wheeler-DeWitt equation and 
obtain its solutions. In sect. 7 we offer some concluding remarks.  
 
\section {Field equations} 
We write the two-dimensional action for $f(R)$ gravity as 
\begin{equation}\label{eq:6} 
I = I_{G} + I_{M}~, 
\end{equation} 
where 
\begin{equation}\label{eq:7} 
I_G = -\frac{1}{2 g_{_N}} \int { d}^2 { x} \sqrt{-g}~ f(R)~, 
\end{equation} 
is the gravitational action  and $I_M$ is the matter action [13]. The field equations can be derived by varying the action with respect to the metric tensor $g_{\mu\nu}$. Upon noting that the stress-energy tensor is defined by 
 
\begin{equation}\label{eq:8} 
\delta I_{M} = \frac{1}{2} \int { d}^2 { x} \sqrt{-g}~T^{\mu\nu} \delta g_{\mu\nu}~, 
\end{equation} 
we derive the following field equation 
 
\begin{equation}\label{eq:9} 
f'(R)~ R_{\mu\nu} -\frac{1}{2} g_{\mu\nu}f(R) -g_{\mu\nu} \Box f'(R) + 
\nabla_\mu \nabla_\mu f'(R) = -g_{_N} T_{\mu\nu}~. 
\end{equation} 
In eq.\eqref{eq:4} $R_{\mu\nu}$ is the Ricci tensor, the prime denotes the differentiation with respect to $R$ and the operator $\Box$ is defined by

\begin{equation} \label{eq:10} 
\Box f'(R) = \frac{1}{\sqrt{-g}}~\partial_\mu (\sqrt{-g}~ g^{\mu\nu} 
~\partial_\nu f'(R))~. 
\end{equation} 
Using eq.\eqref{eq:5} we can write eq.\eqref{eq:9} as 
\begin{equation} \label{eq:11} 
\frac{1}{2} g_{\mu\nu} (f'(R)R -f(R)) - g_{\mu\nu} \Box f'(R) + \nabla{_\mu} 
\nabla_{\nu} f'(R) = -g_{_N} T_{\mu\nu}~. 
\end{equation} 
In the following we shall be concerned with cosmological implicatons of eq.\eqref{eq:11}. For this purpose we shall adopt the Friedman-Robertson-Walker (FRW) metric which in two-dimensional spacetime reads ( $c=1$) 
\begin{equation} \label{eq:12} 
{\rm ds}^2 = -{dt}^2 + \frac{a^2 (t)}{1-k~ x^{2}}~ { dx}^2~, 
 \end{equation} 
in terms of the comoving coordinates $x$ and $t$. The quantity $a(t)$ is the 
usual time-dependent cosmic scale factor. A change of variable ${dx}^2/(1-kx^2)\rightarrow { dx^2}$ leads 
\begin{equation} \label{eq:13} 
{ds}^2 = -{ dt}^2 + a^2( t)~ { dx}^2~. 
\end{equation} 
Thus in two dimensions the time evolution of $a(t)$ is not affected by the value of $k = 0 , \pm 1$ corresponding to the three different cosmological models [14]. This is unlike the four-dimensional case. The values $k = 0, -1$ still describe spatially open flat and hyperbolic universes respectively while $k=1$ describes a closed universe. The stress-energy tensor of the homogeneous isotropic universe is taken to be that of a perfect fluid 
\begin{equation} \label{eq:14} 
T_{\mu\nu} = p~ g_{\mu\nu} +~(p+\rho)~ U_\mu~ U_\nu~, 
\end{equation} 
where $p$ is the pressure, $\rho$ is the energy density and $U^\mu$ is the comoving velocity.  Using eq.\eqref {eq:13} and eq.\eqref {eq:14} we obtain from eq.\eqref {eq:11}  the following two independent cosmological field equations 
  \begin{equation} \label{eq:15} 
\frac{1}{2} (R~f'(R) - f(R)) +\frac{\dot a}{a} \partial_{t} f'(R) = g_{_N}\rho~, 
\end{equation} 
 
\begin{equation} \label{eq:16} 
\frac{1}{2} (R~f'(R) - f(R)) +  \partial_t^2 f'(R) = - g_{_N}~p~, 
\end{equation} 
where we use the dot as well as $\partial_t$ to indicate differentiation with respect to time.  We note that if $f(R)$ is expressed as a sum of powers $R^n$ of $R$, then a term linear in $R$ would cancel out in the bracketed terms in eq.\eqref {eq:15} and eq.\eqref{eq:16} and would not contribute to the derivative terms either.  Hence it has no effect on the dynamics. The stress-energy tensor obeys the conservation law 
 
\begin{equation} \label{eq:17} 
\nabla^\alpha T_{\alpha_\beta} = 0~, 
\end{equation} 
and this, for a perfect fluid, gives rise to the following two equations 
 
\begin{equation}\label{eq:18} 
U^\alpha~\nabla_\alpha \rho +~ (p+\rho)~\nabla^\alpha~U_\alpha =0~, 
\end{equation} 
and 
\begin{equation}\label{eq:19} 
(p+\rho)~U^\alpha~\nabla_\alpha~U_\beta + ~(g_{\alpha\beta} 
  +~U_\alpha~U_\beta)~\nabla^\alpha~p =0 ~. 
\end{equation} 
For the FRW metric of eq.\eqref {eq:13}  one readily obtains from eq.\eqref {eq:18} that 
\begin{equation}\label{eq:20} 
\frac{\rm d}{\rm d \rm a}(\rho a) = -p ~. 
\end{equation} 
Assuming an equation of state of the form $p = \gamma \rho$ where $\gamma$ is a constant, eq.\eqref{eq:20}  immediately leads to 
 
\begin{equation}\label{eq:21} 
\rho = C a^{-\gamma-1}~, 
\end{equation} 
where  $C$ is a constant.  Eq.\eqref{eq:19}  is seen to be identically satisfied and does not give rise to anything new.  For a pressureless (dust) pure matter universe ($\rho_m \neq 0$, $\rho_r = 0$, $\gamma=0$) we have  
\begin{equation}\label{eq:22} 
\rho_m = C_m~ a^{-1}~, 
\end{equation} 
  while for a pure radiation universe ($\rho_m =0,~ \rho_r \neq 0, ~\gamma=1$) one has 
 
\begin{equation}\label{eq:23} 
\rho_r = C_r~ a^{-2}~~, 
\end{equation} 
  Denoting the present time by $t_0$ and using the usual notation of $a_0 
  \equiv a(t_0)$ and $ \rho_0 \equiv \rho(t_0)$ to denote present-day values of these quantities, we can write for a matter-dominated universe 
 
\begin{equation}\label{eq:24} 
p_m = 0,~ \rho_m (t) = \rho_{m0}~ \frac{a_{_0}}{a(t)}~, 
\end{equation} 
while for a radiation-dominated universe one has 
 
\begin{equation}\label{eq:25} 
p_r = \rho_r (t) = \rho_{r0}\bigg(\frac{a_{_0}}{a(t)}\bigg)^2~, 
\end{equation} 
  Finally we wish to note that for the FRW metric the curvature scalar of this two-dimensional universe is given by 
 
\begin{equation}\label{eq:26} 
R = - \frac{2\ddot{a}}{a(t)}~, 
\end{equation} 
where $\ddot a =\frac{\rm d^2{\mathit a}}{\rm d \rm t^2}$. 
 
\section { Solutions of the cosmological field equations} 
In this section we seek solutions of the cosmological field eqs.\eqref{eq:15} 
and \eqref {eq:16} with the energy density and pressure given by 
eqs.\eqref{eq:24} and \eqref{eq:25} for each component of the cosmological fluid thus obtaining two sets of equations. For the matter dominated epoch we obtain the following equations

\begin{equation} \label{eq:27} \frac{1}{2}~(Rf'(R)-f(R) ) + \frac{\dot a}{a} \partial_t f'(R) = g_{_N} \rho_{m0} \frac{a_0}{a}~,\end{equation} 
\begin{equation} \label{eq:28} \frac{1}{2}~(Rf'(R) - f(R)) + \partial^2_t f'(R) = 0~. \end{equation} 
 
  For the radiation dominated epoch the corresponding equations read 
 
\begin{equation} \label{eq:29} \frac{1}{2}~(R f'(R) - f(R)) + \frac{\dot a}{a} \partial_t f'(R) = g_{_N}\rho_{r 0} \frac{a_0^2}{a^2}~, \end{equation} 
\begin{equation} \label{eq:30} \frac{1}{2}~(R f'(R) - f(R)) +  \partial_t^2 f'(R) = -g_{_N}\rho_{r 0} \frac{a_0^2}{a^2}~, \end{equation} 
  To proceed further we need to specify the function $f(R)$.  Similar to the procedure followed in the four-dimensional case [5] we take for $f(R)$ the following expression

\begin{equation} \label{eq:31} f(R) = R + \alpha R^n~, \end{equation} 
  where the real constants $\alpha$ and $n$ are, at this stage, only 
  restricted by $\alpha \neq 0$ and $n \neq 1$.  Upon substitution of eq.\eqref{eq:31} into eqs.\eqref{eq:27} and \eqref{eq:28} we obtain the equations

\begin{equation} \label {eq:32} \frac{1}{2} (n-1) \alpha R^n + n(n-1) \alpha R^{n-2} \dot R \frac{\dot a}{a} = g_{_N} \rho_{m0} \frac{a_0}{a}~, \end{equation} 
\begin{equation} \label{eq:33} 2nR\ddot R + 2n(n-2) \dot R^2 + R^3 = 0 ~. \end{equation} 
Equations \eqref{eq:32} and \eqref{eq:33}  describe the matter dominated epoch and we shall attempt to find solutions for them now.  We start with eq.\eqref{eq:33}  and note that in terms of the function $z(R)$ defined by 
 
\begin{equation} \label{eq:34} z(R) = \dot R^2 ~, \end{equation} 
the equation is transformed into the following form 
 
\begin{equation} \label{eq:35}\frac {{\rm d}{z}}{\rm dR} + \frac{2(n-2)}{R} z + \frac {R^2}{n} = 0~.\end{equation} 
This equation is easily solved and  we obtain for $n \neq \frac{1}{2}$ 
 
\begin{equation} \label{eq:36} z(R) = \dot R^2 = -~ \frac{1}{n(2n-1)} R^3 + C_1 R^{4-2n}~, \end{equation} 
where $C_1$ is a constant.   Eq.\eqref {eq:36} then leads to the parametric solution 
\begin{equation} \label{eq:37} t = \pm \int \bigg[-\frac{R^3}{n(2n-1)} + C_1 R^{4-2n}\bigg]^{-1/2} {\rm d}{R} + C_2~,\end{equation} 
  where $C_2$  is a constant.  For $n = 2$ and  $C_1 \neq 0$ one can carry out the integration using the result [14] 
 
\begin{equation}\label{eq:38}  
\int \frac{ dx}{{(K-x^{\alpha+2})}^{1/2}}= 
  \frac{x}{\sqrt K}~ {_2F_1} \bigg( \frac{1}{2}, \frac{1}{\alpha+2}, 
  \frac{\alpha+3}{\alpha+2} ; \frac {x^{\alpha+2}}{K}\bigg)~,  
\end{equation} 
 
where $\alpha, K$ are constants and $_2{F}_1 $  is the hypergeometric function.  We obtain 
 
\begin{equation} \label{eq:39}t = \pm \bigg( \frac{6}{C_1}\bigg)^{1/2} R~ {_2F_1} \bigg( \frac{1}{2}, \frac{1}{3}, \frac{4}{3}; \frac{R^3}{C_1}\bigg) + C_2~. \end{equation} 
Ideally one should solve eq.\eqref{eq:39} to obtain $R$ as a function of the 
cosmic time $t$ and plug that into eq.\eqref {eq:32} in order to solve for $a(t)$ in the case of $n = 2$, but that is a difficult task.  Instead we consider solutions for which $C_1= 0$ in eq.\eqref{eq:36} and a general $n \neq \frac{1}{2}$.  One can then easily derive that 
 
\begin{equation} \label{eq:40} R = -~\frac{4n(2n-1)}{(t-t_m)^2}~, \end{equation} 
where we have renamed the integration constant $C_2$ as $t_m$.  In fact one 
can verify directly by substitution that the expression for $R$ in 
eq.\eqref{eq:40} is a solution of eq.\eqref{eq:33}  
\vglue0.1cm\hglue-0.6cm 
Next we substitute eq.\eqref{eq:40} into eq.\eqref{eq:32} and obtain 
 
\begin{equation} \label{eq:41} A_2 \dot a + A_1 (t-t_m)^{-1} a = K a_0~ (t-t_m)^{2n-1}~, \end{equation} 
where 
\begin{equation} \label{eq:42} A_1 = \frac{1}{2}(n-1) N^n;~~ A_2 = -2n(n-1)N^{n-1}~, \end{equation} 
and 
\begin{equation} \label{eq:43} N = 4n (1-2n); ~~ K = \frac{g_{_N} \rho_{m0}}{\alpha}~. \end{equation} 
We readily solve eq.\eqref{eq:41}  and get 
\begin{equation} \label{eq:44} a(t)= \bar{C} (t-t_m)^{1-2n} + \bar{K} 
  (t-t_m)^{2n}~, \end{equation} 
where $\bar C$ is a constant and 
\begin{equation} \label{eq:45} \bar{K} = \frac{K a_0}{(4n-1) A_2} ~. \end{equation} 
Clearly $n$ must be such that $A_1$ and $A_2$ are real and $\bar K$ is finite. 
We will return to this issue later.  It is interesting to note that the $t$ 
dependence of $R$ is $(t-t_m)^{-2}$ and  thus independent of $n$, while that 
of $a(t)$ does depend on $n$.  We also note that the relation $R = - 
\frac{2\ddot a}{a}$ is satisfied by the solutions for $R$ and are given in eqs.\eqref{eq:40} and \eqref{eq:44} respectively. We further note that the second term in eq.\eqref {eq:44}  is a solution of eq.\eqref{eq:41} in its own right.   On the other hand the first term in eq.\eqref{eq:44}  is a solution of the homogeneous form of eq.\eqref{eq:41}.  Furthermore the constants $\bar C$ and $\bar K$ must be such that $a(t)$  is positive.\\    
We now turn to the case of radiation.  Upon adding eqs.\eqref{eq:29}  and \eqref{eq:30}  we obtain the equation 
 
\begin{equation} \label{eq:46} Rf'(R) - f(R) + \partial_t^2 f'(R) + \frac{\dot a}{a} \partial_t f'(R) = 0~. \end{equation} 
Employing in eq.\eqref{eq:46}  the expression for $f(R)$ given in eq.\eqref{eq:31}  above yields

\begin{equation} \label{eq:47} n R \ddot R + n(n-2) \dot R^2 + R^3 + nR \dot R \frac{\dot a}{a} = 0~.\end{equation} 
  Next we use eq.\eqref{eq:31} in eq.\eqref{eq:29}  and obtain 
 
\begin{equation} \label{eq:48} \frac{1}{2} (n-1) \alpha R^n + n(n-1) \alpha R^{n-2} \dot R \frac{\dot a}{a} = g_{_N}\rho_{r0} \frac{a_0^2}{a^2}~. \end{equation} 
Motivated by the structure of the solutions for the cosmological equations in 
the case of pure matter above, we seek solutions for $R(t)$ and $a(t)$ of eqs.\eqref{eq:47} and \eqref{eq:48} in the form of powers in $t-t_r$ where $t_r$ is some reference time.  We obtain the following results 
 
\begin{equation} \label{eq:49} R(t) = 2n (n-1)(t-t_r)^{-2}~, \end{equation} 
 
\begin{equation}\label{eq:50} a(t) = B (t-t_r)^n ~, \end{equation} 
  where the constant $B$ is given by 
 
\begin{equation} \label{eq:51} B = \bigg[ \frac{g_{_N} \rho_{r0}}{n(n-1)(1-3n)[2n(1-n)]^{n-1} \alpha}\bigg]^{1/2} a_0~. \end{equation} 
Note that, as in the case of matter, the $t$ dependence of $R(t)$ is 
independent of $n$, the only such dependence appears in the overall 
coefficient.  We also note that the relation $R = -\frac{2 \ddot a}{a}$ is 
satisfied by the solutions for $R$  and are given above in eqs.\eqref{eq:48} 
and \eqref{eq:50}.  For an expanding universe one must have $n > 1$ and  $B > 0$.  Furthermore the value of $n$ must ensure that the bracketed term in eq.\eqref{eq:51} is finite and real. 
 
\section{Properties of the solutions} 
We now discuss some properties of the solutions of the cosmological field equations found in the previous section.  Let us first look at the radiation dominated case and determine whether our vision of the universe is limited by a particle horizon.  At a given cosmic time $t_s$ the proper distance {\rm d(t$_s$)} of the emitter is given by 
\begin{equation} \label{eq:52} { d} ({ t_s}) = a (t_s) \int^{t_s}_{t_e} \frac{{\rm d}t'}{a(t')}~, \end{equation} 
Where $t_e$  is the time of emission of the photon.  Using eq.\eqref{eq:50} we obtain 
\begin{equation} \label{eq:53} { d} ({ t_s}) = \frac{(t_s -t_r)^n}{1-n} \bigg[(t_s-t_r)^{1-n} - (t_e -t_r)^{1-n}\bigg]~.\end{equation} 
We can view $t_r$ as signifying the onset of the radiation epoch.  We see that 
as $t_e \rightarrow t_r$,  {\rm d(t$_s$)} is finite for $1-n > 0$  and 
diverges for $1-n < 0$.  Hence no particle horizon problem will arise if $n > 
1$ which is the same condition required for an expanding universe.  Reality of 
$B$ also requires $n$ to be an integer.  For $n$ an even integer, the 
parameter $\alpha$  must be positive while for $n$ odd, $\alpha$ should be 
negative.  Thus we take $n$ to be a positive integer greater than one. Next we note that the cosmic acceleration $\ddot a (t)$  which is given by 
 
\begin{equation}\label{eq:54} \ddot a (t) = n(n-1)~ B~ (t-t_r)^{n-2}~, 
\end{equation}  
is positive for $ t> t_r$ since $n>1$  and is constant for $n = 2$.  Now in 
two-dimensional spacetime the radiation energy density $\rho_r \propto T^2$ 
were $T$ is the temperature [14] and it follows therefore from eq. \eqref 
{eq:25} that    
\begin{equation} \label{eq:55} a \propto T^{-1} ~. \end{equation} 
Since we have $a \rightarrow 0$  as $t \rightarrow t_r$, we conclude that this radiation universe has a hot big bang origin.\\ 
Next we turn to the case of the matter dominated universe described by eqs. \eqref{eq:40}  and \eqref{eq:44} .  First let us consider the case $\bar C =0$  when the scale factor becomes 
\begin{equation} \label{eq:56} \bar{a}(t) = \bar{K} 
  (t-t_m)^{2n}~. \end{equation}  
As we have stated earlier this is viable because it represents a solution of eq.\eqref{eq:41}.  The time $t_m$ can be taken to signify the onset of matter dominance.  The proper distance {\rm d(t$_s$)} is now given by 
\begin{equation} \label{eq:57} { d}({ t_s}) = \frac{(t_s - t_m)^{2n}}{1-2n} [(t_s - t_m)^{1-2n} - (t_e -t_m)^{1-2n}]~. \end{equation} 
Hence no particle horizon will arise if $2n > 1$.  Also as we stated following 
eq.\eqref{eq:45}, the parameter $n$ must be such that the constants $A_1$ and 
$A_2$  given by eq.\eqref{eq:42} are real.  Since for $2n > 1$ the number $N$ 
of eq.\eqref{eq:43} is negative, it follows that $n$ has to be a positive 
integer.  Now the requirement that $\bar a(t) > 0$ for $t > t_m$ implies that 
$\bar K> 0$.  For $n$ even we have  $A_2 > 0$ and hence $\alpha$  should be 
positive to ensure $\bar K> 0$ while for $n$ odd one has $A_2 < 0$ and $\alpha$ 
should be negative.  Since we exclude $n = 1$, the smallest permissible value 
is $n = 2$.  For such values of $n$ it is evident that the cosmic acceleration 
$\ddot {\bar a}(t)$ is positive.  Finally we observe that for the pure matter 
universe we have  $\bar a(t) \rightarrow 0$ as  $t \rightarrow t_m$.\\ 
We now consider the case $C \neq 0$.  Using eq.\eqref{eq:44}  the proper distance is now given by 
 
\begin{equation} \label{eq:58}{\rm d}({\rm t_s}) = a(t_s) \int^{t_s}_{t_e} \frac{(t-t_m)^{2n-1}}{C+\bar{K} (t-t_m)^{4n-1}} {\rm d}t~. \end{equation} 
It is clear that the integral converges for $t_e \rightarrow t_m$ and we do a have particle horizon.  Performing the integral we determine the proper distance to the horizon to be 
 
\begin{equation*}\label{eq:59} {\rm d}({\rm t_s}) = \frac{a(t_s)}{\bar C} \bigg(\frac{\bar K}{\bar C}\bigg)^{2n(4n-1)}\bigg\{- \frac{ln(1+\xi_s)}{4n-1}\end{equation*} 
 
 \begin{equation*}-\frac{1}{4n-1} \sum_{k=1}^{2n-1} cos \bigg[\frac{2n\pi(2k-1)}{4n-1}\bigg] ln\bigg(1-2 \xi_s~ cos \frac{2k-1}{4n-1} \pi + \xi_s^2 \bigg)  \end{equation*} 
 
\begin{equation} +\frac{2}{4n-1} \sum_{k=1}^{2n-1} sin \bigg[\frac{2n\pi (2k-1)}{4n-1}\bigg] arctg \bigg[\frac {\xi_s - cos \frac{2k-1}{4n-1} \pi}{sin \frac{2k-1}{4n-1} \pi }\bigg] - (\xi_s \longleftrightarrow \xi_e)\bigg\}~~. \end{equation} 
where 
\begin{equation} \label{eq:60}\xi_j = \bigg( \frac{\bar C}{\bar K}\bigg)^{4n-1} ( t_j - t_m), ~ j = s,e~. \end{equation} 
Let us now study further properties of the solution given in eq.\eqref{eq:44}. 
In the following we consider only values of $t$  such that $t>t_m$ .  Now it 
is evident that, except for values of $n$ in the interval $0 < n < 
\frac{1}{2}$,  the first term in eq.\eqref{eq:44} dominates for $t$ near $t_m$ 
when $n > \frac{1}{2}$  while the second term dominates for $n<0$.  Hence to 
ensure positivity of the scale factor we require that both $\bar C$ and $\bar 
K$ be positive.  For $0 < n < \frac{1}{2}$ , $\bar C$ and $\bar K$ can have 
opposite signs but only in such a manner so as to keep $a > 0$. We shall for 
simplicity assume that $\bar C>0$ and $\bar K > 0$  for all values of $n$.  Next we observe that outside the interval $0 < n < \frac{1}{2}$ , the number $N$ of eq.\eqref{eq:43} is negative and  to ensure the reality of $A_2$  given by eq.\eqref{eq:42}, the number $n$ has to be an integer.  We readily deduce that for $\alpha > 0$, $n$  can be a positive even integer or a negative odd integer.  On the other hand for $\alpha < 0$, $n$ can be a positive odd integer or a negative even integer.  The cosmic acceleration $\ddot a(t)$ is given by 
\begin{equation} \label{eq:61} \ddot a(t) = 2n (2n-1) (t-t_m)^{-2} a(t)~. \end{equation} 
It is seen that $\ddot a < 0$  for $0 < n < \frac{1}{2}$, $\ddot a = 0$ for $n 
= \frac{1}{2}$ and  $\ddot a > 0$ for $n < 0$ or~ $n >\frac{1}{2}$ . 
\vglue0.1cm\hglue-0.65cm 
Next we consider the behavior of $a(t)$ as $t \rightarrow t_m$ for the case 
$\bar C \neq 0$.  We see from eq.\eqref{eq:44}  that for  $0 < n < 
\frac{1}{2}$, $a(t) \rightarrow 0$ as $t \rightarrow t_m$ and 
accordingly the temperature $T \rightarrow \infty$  in this limit.  For $n 
= \frac{1}{2}$, we have $a(t) \rightarrow \bar C$  as $t \rightarrow 
t_m$ and $T$  is finite.  However for $n$ outside the interval  $0\leq n \leq 
\frac{1}{2}$  the behavior of $a(t)$ is very different as $t \rightarrow 
t_m$.   We see that $a(t) \rightarrow \infty$  in this limit and and 
energy density $\rho_m$ as well as the temperature tend to zero.  As $t$  
increases beyond the value $t_m$, $a(t)$ decreases to finite values and the 
density increases.  However $a(t)$ never reaches zero and attains a minimum 
value at $t = t_c$  given by   
 
\begin{equation} \label{eq:62} t_c = t_m + \bigg[\frac{(2n-1) \bar C}{2n \bar K}\bigg]^{1/{4n-1}}~. \end{equation} 
For $t>t_c$, $a(t)$ starts to increase.  We also note from eq.\eqref{eq:40} that the curvature scalar  $R \rightarrow -\infty$  as $t \rightarrow t_m$ and then starts increasing through finite negative values as $t$ grows beyond $t_m$.  The singular behavior of the scale factor noted here should be contrasted with that of the FRW cosmological models in four-dimensional general relativity where the scale factor and energy density go to zero and infinity respectively as the initial moment is approached.   
 
\section{Inflation} 
The horizon problem in four-dimensional standard FRW cosmology is a consequence of deceleration in the expansion of the universe.  The problem can be solved by postulating a phase of the universe, prior to the decelerating phase, in which the expansion is accelerating and such a phase is called a period of inflation.   Hence inflation is characterized by the following property for the scale factor $a(t):$ 
\begin{equation}\label{eq:63} \ddot a(t)  > 0 ~.\end{equation} 
Now as evident from the analysis of sect.  4, $\ddot a > 0$ is readily achieved in our $f(R)$ theory in two-dimensional spacetime and the universe is accelerating.  The solutions obtained for the scale factor displayed power dependence on time akin to that of power-law inflation.  It would seem that there is no need to require an inflationary phase since matter or radiation dominated epochs yield an accelerating universe.  Here we are not seeking to introduce scalar fields to propel acceleration as in the usual inflationary cosmology.  We recall that one of the motivations for introducing modified or $f(R)$ theories of gravity in four-dimensional spacetime is the desire to explain acceleration of the universe as an alternative to using scalar fields.  For this purpose solutions for the cosmological field equations are sought in the absence of the matter fluid [5].  We carry out such an analysis in our case by considering solutions to eqs.\eqref{eq:32}  and \eqref{eq:33}  of sect.  3 with the R.H.S set equal to zero.  We have earlier obtained a general solution for eq.\eqref{eq:33}  given by eq.\eqref{eq:37} of sect.  3.  However the parametric nature of that solution makes it difficult to use in eq.\eqref{eq:32}  in order to solve for $a(t)$.  Putting $C_1 = 0$ enables the integration in eq.\eqref{eq:37} to be performed and leads to the solution given in eq.\eqref{eq:40}  which we write below as  
 
\begin{equation}\label{eq:64} R = - \frac{4n(2n-1)}{(t-\bar t)^2}~, \end{equation} 
where $n \neq \frac{1}{2}$ , 1 and we have now denoted the integration constant by $\bar t$.  Using eq.\eqref{eq:64}  in eq.\eqref{eq:32} with the R.H.S. set equal to zero yields the equation 
 
\begin{equation}\label{eq:65} \dot a + \frac{2n-1}{t-\bar t}~a = 0~, \end{equation} 
the solution of which reads 
 
\begin{equation}\label{eq:66} a(t) = A ( t- \bar t)^{1-2n}~, \end{equation} 
where $A > 0$ is a constant.  We take the solution to hold for $t > \bar t$ .  The cosmic acceleration is given by 
 
\begin{equation}\label{eq:67} \ddot a(t) = 2n(2n-1)~A~(t-\bar 
  t)^{-2n-1}~.\end{equation}  
The Hubble parameter is 
\begin{equation}\label{eq:68}  H = \frac{\dot a}{a} = \frac{1-2n}{t-\bar t} ~,\end{equation} 
and 
\begin{equation}\label{eq:69} \dot H = \frac{2n-1}{(t-\bar t)^2}~.          \end{equation} 
For  $n < -1$ we can identify $\bar t$ with the onset of inflation $\bar t = t_i$.  Eq.\eqref{eq:66} then describes a universe that expands with positive acceleration for $t > t_i$ .  We also have $H > 0$ and $\dot H < 0$ for $ t > t_i$ which characterizes standard inflation.  However if we make the identification $\bar t = t_i$ for  $n > 1$, we will have a situation in which $a (t) \rightarrow \infty$  as $t \rightarrow t_i$  thus obtaining a universe that starts off already with an infinite size at the onset of inflation collapsing subsequently for $t > t_i$ at an accelerated rate.  Such a scenario can be avoided if $\bar t$ is instead taken to have a relatively large value so that $t < \bar t$  during the inflationary epoch.  We write $a (t)$ now as 
\begin{equation}\label{eq:70}  a(t) = A ~ |t-\bar t|^{1-2n} ~.         \end{equation} 
The universe then starts off with a relatively small non-zero size at $t =t_i$ and expands with positive acceleration as time progresses.  We also have 
\begin{equation}\label{eq:71}  H = \frac{2n-1}{\bar t -t} ~, 
\end{equation}  
 
\begin{equation}\label{eq:72}       \dot H = \frac{1-2n}{(\bar t- t)^2}~,     \end{equation} 
so that $H > 0$  and $\dot H < 0$  and we again have standard inflation.\\ 
As in four spacetime dimensions we define the so called slow-roll parameter $\varepsilon$ by [5] 
\begin{equation} \label{eq:73} \varepsilon = -\dot H /H^2 ~.            \end{equation} 
and  in terms of which one has 
\begin{equation} \label{eq:74} \frac{\ddot a}{a} = H^2 + \dot H = (1-\varepsilon) H^2~.            \end{equation} 
Inflation can thus be attained only if $\varepsilon < 1$.  In our present context $\varepsilon$ is given by 
\begin{equation} \label{eq:75}\varepsilon= \frac{1}{|2n-1|}~,       \end{equation} 
for both cases of $n < -1$ and $n > 1$  we clearly have $\varepsilon < 1$. 
The slow-roll approximation corresponding to $\varepsilon <<1$, then obtains 
when $|2n-1|>> 1$. 
\vglue0.1cm\hglue-0.65cm 
As we have stated above the solution for $R$ given in eq.\eqref {eq:64}  arises as a special case of the general solution given in eq.\eqref{eq:37}.  As an alternative to solving eqs.\eqref{eq:32} and \eqref{eq:33} one can derive an equation for the Hubble parameter [15], [5].  We write eq.\eqref{eq:32}with the R.H.S set equal to zero 
\begin{equation} \label{eq:76} 2n \dot R \dot a + R^2 a=0 ~.                               \end{equation} 
Now from eq.\eqref{eq:26}  of sect.  2 we obtain 
\begin{equation} \label{eq:77}  \dot R = -\frac{2 \dddot a}{a} +\frac {2 \ddot a ~\dot a}{a^2}~.  \end{equation} 
Substituting eqs.\eqref{eq:26} and \eqref{eq:77} in eq.\eqref{eq:76} one obtains 
\begin{equation} \label{eq:78} - n~ a~ \dot a~ \dddot a + a~ \ddot a^2 + n \dot a^2 \ddot a =0 ~.    \end{equation} 
Next in terms of $H$, $\dot H$ and $\ddot H$ we can express eq.\eqref{eq:78}, after some manipulations, as 
\begin{equation} \label{eq:79} -nH \ddot H -2(n-1) \dot H H^2 + \dot H^2 + H^4 = 0  ~. \end{equation} 
It is customary, in dealing with equations such as this, to invoke the 
slow-roll approximation $|\dot H/H^2|<<1$ and  $|\ddot H/H \dot H|<<1$,  [15], [5].  Applying this to eq.\eqref{eq:70} we obtain that 
\begin{equation} \label{eq:80}  -2(n-1) \dot H + H^2 = 0  ~.                              \end{equation} 
The solution of eq.\eqref{eq:80}  is 
\begin{equation} \label{eq:81} H(t) = \frac{-2(n-1)}{t- \bar {t}~^{'}}~,                                \end{equation} 
where $\bar{t}'$  is a constant.  Eq.\eqref{eq:81} in turn gives 
\begin{equation} \label{eq:82}  a(t) = A'/(t-\bar{t}~^{'})^{2(n-1)}~.                               \end{equation} 
with $A'$  being another constant.  Eq.\eqref{eq:82} for $a(t)$ is similar in 
structure to eq.\eqref{eq:70} and the properties of the solution are 
therefore similar to what we discussed before and hence will not be 
considered any further.\\ 
We shall next seek a general solution to eq.\eqref{eq:33} for $R (t)$ that 
holds for  
$t$ close to the instant $t_i$  that signifies the onset of inflation. 
Specifically we assume that $t=t_i$ is a regular point of eq. \eqref{eq:33} 
and seek a solution for $R(t)$  in the form of  a power series confining 
ourselves to small values of~ $t-t_i$.   For simplicity we consider the case $n 
= 2$  for which eq. \eqref{eq:33} becomes  
\begin{equation} \label{eq:83}  4\ddot R +R^2=0 ~.                              \end{equation} 
We write 
\begin{equation}\label{eq:84} R(t)= \sum_{m=0}^{\infty}~ b_m (t-t_i)^m~.    \end{equation} 
Substituting eq. \eqref{eq:84}  in eq.\eqref{eq:83}  and solving we obtain 
\begin{equation}\label{eq:85} b_2 = - \frac{1}{8} b_0^2~,    \end{equation} 
 
\begin{equation}\label{eq:86} b_3 = -~\frac{1}{12} b_0~b_1~,    \end{equation} 
etc.  This leads to 
\begin{equation}\label{eq:87}  R(t) = b_0 + b_1 (t-t_i) - \frac{1}{8} b_0^2 (t-t_i)^2 - \frac{1}{12}b_0 b_1(t-t_i)^3 +......   \end{equation} 
We remark that if inflation lasts for a short period of time then it is sensible to have a representation for $R(t)$ as given in eq.\eqref{eq:87}.  Moreover for sufficiently small $t-t_i$ we can approximate $R(t)$ by the first two terms and substitute in eq.\eqref{eq:76} with $n=2$.  Solving the resulting equation we obtain 
\begin{equation}\label{eq:88}a(t)\approx C~exp\bigg\{-\frac{1}{12b_1^2}[b_0 + b_1 (t-t_i)]^3\bigg\}~,     \end{equation} 
where $C > 0$  is a constant.  We can write eq.\eqref{eq:88}  as 
\begin{equation}\label{eq:89}  a(t)  \approx 
  a_i~exp\bigg\{-\frac{1}{12b_1^2}~([b_0 + b_1 (t-t_i)]^3- b_0^3)\bigg\} ~, 
\end{equation} 
where 
\begin{equation}\label{eq:90}  a_i = a(t_i) = C~exp\bigg( -~\frac{b_0^3}{12b_1^2}\bigg) ~.  \end{equation} 
From eq. \eqref{eq:88} we obtain 
\begin{equation}\label{eq:91} \dot a (t) = -~\frac{1}{4b_1} [b_0 + b_1(t-t_i)]^2 ~a(t) ~,    \end{equation} 
and 
\begin{equation}\label{eq:92}\ddot a (t)=\bigg\{ -~\frac{1}{2}[b_0+b_1(t-t_i)]+\frac{1}{16b_1^2} [b_0 +b_1(t-t_i)]^4\bigg\} a(t)~.     \end{equation} 
From eq. \eqref{eq:89} we see that we must have $b_1< 0$  to ensure that $\dot a > 0$.  We must also require $a(t)$ to be increasing for  $t>t_i$.   This can be achieved by having $b_0 > 0$ for then $b_0 + b_1 (t- t_i)$  will start off at the value $b_0$ and decreases reaching zero at $t^* - t_i = - \frac{b_0}{b_1}$.  During the interval, $t_i < t < t^*$,  $a(t)$  will be increasing.  We must also require the cosmic acceleration $\ddot a(t)$ to be positive during this interval and this leads to the condition 
 
\begin{equation}\label{eq:93}  \frac{1}{8b_1^2} [b_0+b_1(t-t_i)]^3 > 1~.   \end{equation} 
This inequality will continue to hold until $t=t_f < t^*$  when $\ddot a (t_f)= 0$.  This implies that 
\begin{equation}\label{eq:94} \frac{1}{8b_1^2}~[b_0 +b_1(t_f-t_i)]^3 =1 ~,   \end{equation} 
which yields 
\begin{equation}\label{eq:95} t_f  = t_i + b_0 |b_1|^{-1} - 2|b_1|^{-1/3}~.   \end{equation} 
The time $t_f$  then signifies the end of inflation.  Since $R(t_i) = b_0$ 
and  $\dot R(t_i) =b_1$, the conditions $b_0 > 0$  and   $b_1 < 0$  can be 
expressed as  
\begin{equation}\label{eq:96}  R(t_i) > 0    \end{equation} 
\begin{equation}\label{eq:97} \dot R (t_i) < 0~.    \end{equation} 
We can also express the duration of inflation as  
\begin{equation}\label{eq:98} t_f - t_i = R(t_i) |\dot R(t_i)|^{-1} - 2 |\dot R (t_i)|^{-1/3}~.    \end{equation} 
The Hubble parameter is given by 
\begin{equation}\label{eq:99}H = -~\frac{1}{4b_1} [b_0 + b_1(t-t_1)]^2 ~. 
\end{equation}  
It thus decreases from an initial value $ H_i$  given by 
 
\begin{equation}\label{eq:100} H_i = H(t_i) = -~\frac{b_0^2}{4b_1} = \frac{R^2(t_i)}{4|\dot R(t_i)|}~,    \end{equation} 
to a value $H_f$ at the end of inflation where 
\begin{equation}\label{eq:101}  H_f = H(t_f) = |b_1|^{1/3} = |\dot R (t_i)|^{1/3}~.   \end{equation} 
We note that 
\begin{equation}\label{eq:102} \dot H = - ~\frac{1}{2}~ [b_0 + b_1(t-t_i)]~,    \end{equation} 
is negative during $t_i < t < t_f$  and  we thus have standard inflation.  The slow-roll parameter is given by 
\begin{equation}\label{eq:103} \varepsilon = 8~b_1^2 [b_0 + b_1(t-t_i)]^{-3}~.    \end{equation} 
We recall that for inflation to proceed one must have $\varepsilon < 1$  and 
this leads to precisely the condition expressed in eq.\eqref{eq:93}  stated 
earlier. 
\vglue0.1cm\hglue-0.65cm 
The number of $e$-foldings from $t=t_i$  to $t=t_f$  is defined by [5], [16] 
\begin{equation}\label{eq:104}N = \int_{t_i}^{t_f} H {\rm dt}~, 
\end{equation}  
which is evaluated to give 
\begin{equation}\label{eq:105}  N = \frac{2}{3}\bigg[(H_i/H_f)^{3/2} - 1\bigg]~.   \end{equation} 
In four dimensions, the solution of the horizon and  flatness problems of big bang cosmology requires that $N\geq 70$ , [17], [5].  If we assume that we can use this value in our two-dimensional universe, we find that 
\begin{equation}\label{eq:106} \frac{H_i}{H_f} \geq 22 ~~.   \end{equation} 
i.e., the Hubble parameter decreases to about $4.5\%$ of its initial value by 
the time inflation ends.

\section{Quantization} 
As we stated in the introduction two-dimensional spacetime models of gravity 
provide an arena where issues like quantization are studied since in such a 
setting they prove to be more tractable than in four-dimensional spacetime. In 
this section we thus consider quantization of the $f(R)$ gravity theory 
defined by the action of eq.(3). Our objective is to derive the Wheeler-DeWitt 
equation for the wave function of the universe and obtain its solutions. Since 
we are considering a spatially homogeneous and isotropic universe, we drop the 
spatial intergral and write the action as  
\begin{equation} 
I_G = -\frac{1}{2g_{_N}} \int{dt~ a(t)~ f(R(t))}~~. 
\end{equation} 
We take for f(R) the expression given in eq.(31) and put  $n=2$. We use 
eq.(26) that expresses the scalar curvature in terms of the scale factor and 
write  
\begin{equation} 
I_G=-\frac{1}{2g_{_N}}\int dt (a~R-2~\alpha~ \ddot{a}~R)~~. 
\end{equation} 
We notice the appearance of the second derivative of $a$ in eq.(108). The 
standard approach is to express the wave function in terms of $a$ and $R$ 
[19]. Hence integrating by parts in eq.(108), we obtain 
\begin{equation} 
I_G= \int{{\mathscr{L}}(a,\dot{a},k,\dot{k})~dt}~~, 
\end{equation} 
where 
\begin{equation} 
\mathscr{L} =-\frac{1}{2~g_{_N}}(a~R+2~\alpha~ \dot{a} \dot{R})~~. 
\end{equation} 
The canonical momenta are defined in the usual way 
\begin{equation} 
P_a = \frac{\partial{\mathscr{L}}}{\partial{\dot{a}}} = -\frac{\alpha}{g_{_N}} 
    \dot{R}~~, 
\end{equation} 
\begin{equation} 
P_R = \frac{\partial{\mathscr{L}}}{\partial{\dot{R}}} = -\frac{\alpha}{g_{_N}} 
    \dot{a}~~. 
\end{equation} 
The Hamiltonian is then obtained as 
\begin{eqnarray} 
\mathscr{H}& =&  P_a \dot{a} + P_R \dot{R} -\mathscr{L} \nonumber\\ 
&=& -\frac{g_{_N}}{\alpha} P_a P_R +\frac{1}{2g_{_N}} a~R~~. 
\end{eqnarray} 
Replacing $P_a$ and $P_R$ by $-i\frac{\partial}{\partial a}$ and 
$-i\frac{\partial}{\partial R}$ respectively in the Hamiltonian we obtain the 
Wheeler-DeWitt equation for the wave function of the universe 
\begin{equation} 
\Big(\frac{g_{_N}}{\alpha} \frac{\partial^2}{\partial R~ \partial 
  a}+\frac{1}{2g_{_N}}~ a~R\Big)\psi(a,R)=0~~. 
\end{equation} 
Instead of $a$ and $R$ we shall work with the variables 
\begin{equation} 
\xi =R+a,~~~~~~~~~~~\eta=R-a~~. 
\end{equation} 
In terms of $\xi$ and $\eta$ the Wheeler-DeWitt equation becomes 
\begin{equation} 
\Big[ \frac{g_{_N}}{\alpha} \Big(\frac{\partial^2}{\partial 
  \xi^2}-\frac{\partial^2}{\partial \eta^2}\Big)+\frac{1}{g_{_N}}(\xi^2-\eta^2)\Big]\psi(\xi,\eta)=0~~. 
\end{equation} 
We seek solutions of eq.(116) in factorizable form 
\begin{equation} 
\psi(\xi,\eta)= X(\xi)~Y(\eta)~~, 
\end{equation} 
and obtain the following equations for the functions $X$ and $Y$ 
\begin{equation} 
\frac{d^2 X}{d\xi^2} + \frac{\alpha}{8 g_{_N}^2}\xi^2 X 
=\frac{C~\alpha}{g_{_N}}~ X~~, 
\end{equation} 
\begin{equation} 
\frac{d^2 Y}{d\eta^2} + \frac{\alpha}{8 g_{_N}^2}\eta^2~ Y 
=\frac{C~\alpha}{g_{_N}} Y~~, 
\end{equation} 
where C is the separtation constant. The two equations are identical and hence 
it is enough to consider one of them. We first take $\alpha > 0$  and define 
\begin{equation} 
\gamma^2 = \frac{\alpha}{8 g_{_N}^2}~~, 
\end{equation} 
\begin{equation} 
E = -\frac{C~\alpha}{2~g_{_N}}~~. 
\end{equation} 
In terms of $\gamma^2$ and $E$ eq.(118) reads 
\begin{equation} 
\frac{d^2 X}{d\xi^2}+\gamma^2\xi^2 X + 2~E~X=0~~. 
\end{equation} 
It is interesting to note that eq.(122) is identical to that describing the 
inverted or reversed oscillator discussed by several authors in a number of 
contexts[20]-[24]. By performing the change of variable 
\begin{equation} 
y=\sqrt{2\gamma}~ \xi~~, 
\end{equation} 
we cast eq.(122) into the form 
\begin{equation} 
\frac{d^2X}{dy^2}+\frac{1}{4}~y^2~X +\varepsilon~X=0~~, 
\end{equation} 
where $\varepsilon=E/\gamma$. Eq.(124) is one of the standard forms of 
the equation for the parabolic cylinder functions. Two linearly independent 
solutions are given by the real functions $W(\varepsilon, y)$ and 
$(\varepsilon,-y)$ [25]. For $|y|>>1$ and $|y|>>|\varepsilon|$ these 
solutions display the following asymptotic behaviour 
\begin{equation} 
W(\varepsilon,~ y\rightarrow \infty)\sim \sqrt{\frac{2k}{y}}~ \text{cos}\Big(\frac{1}{4} y^2 
+\varepsilon~ln~y + \frac{1}{4}~\pi +\frac{1}{2} \phi\Big)~~, 
\end{equation} 
\begin{equation} 
W(\varepsilon,~ y\rightarrow -\infty)\sim \sqrt{\frac{2}{k~|y|}}~ \text{sin}\Big(\frac{1}{4} y^2 
+\varepsilon~ln|y| + \frac{1}{4}~\pi +\frac{1}{2} \phi\Big)~~, 
\end{equation} 
where  
\begin{equation} 
k=(1+e^{-2\pi\varepsilon})^{1/2} - e^{-\pi\varepsilon}~~, 
\end{equation} 
\begin{equation} 
\phi = arg~ \Gamma\Big(\frac{1}{2} -i \varepsilon\Big)~~. 
\end{equation} 
The functions $W(\varepsilon, y)$ and $W(\varepsilon, -y)$ satisfy the 
following normalization conditions [22]

\begin{equation} 
\int_{-\infty}^{\infty} W(\varepsilon,y)~W(\varepsilon', -y)~dy  
=\left\{\begin{array}{lll} 
0 & \quad \text{if~ $\varepsilon \neq \varepsilon'$}\\ 
\frac{\pi~e^{-\pi \varepsilon}}{(1+e^{-2\pi\varepsilon})^{1/2}} & \quad 
\text{if~ $\varepsilon=\varepsilon'$}~~. 
\end{array}\right. 
\end{equation}

\begin{equation} \int_{-\infty}^{\infty} W(\varepsilon,y)~W(\varepsilon', y)~dy = 2\pi (1+e^{-2\pi\varepsilon})^{1/2} \delta(\varepsilon-\varepsilon')~~. 
\end{equation} 
The parablolic cylinder functions can be expressed in several forms [25] and 
we can use the various relations between these forms to express $W(a,x)$ in 
terms of the more familiar function $D_p(x)$ for some $p$. In fact one can easily derive that 
\begin{equation} 
W(\varepsilon, y) = \Big(\frac{k}{2}\Big)^{1/2}\big[ e^{i\theta}~D_{i\varepsilon-{1/2}}(y~e^{-\frac{i}{4}\pi}) +e^{-i\theta}~D_{-i\varepsilon-{1/2}}(y~e^{\frac{i}{4}\pi}) \Big]~~, 
\end{equation} 
where 
\begin{equation} 
\theta = \frac{1}{2}\Big(-\frac{1}{2}\pi \varepsilon +\frac{i}{4} \pi + i\phi \Big)~~. 
\end{equation} 
Next we observe that the solutions to eq.(119) are identical to those of eq.(118) but expressed in terms of the vriable $\eta$. Hence we can write the following for the wavefunction $\psi$. 
\begin{equation} \psi(\xi, \eta) = \psi_1(\xi) \psi_2(\eta) 
\end{equation} 
where 
\begin{equation} 
\psi_1(\xi) = C_1~W(E/\gamma, \sqrt{2\gamma}~ \xi) + C_2 W(E/\gamma, -\sqrt{2\gamma}~ \xi)~~, 
\end{equation} 
\begin{equation} 
\psi_2(\xi) = C_1'~W(E/\gamma, \sqrt{2\gamma}~ \eta) + C_2' W(E/\gamma, -\sqrt{2\gamma}~ \eta)~~. 
\end{equation} 
We now consider the case in which the parameter $\alpha$ is negtive and write eqs. (118) and (119) as 
\begin{equation} 
\frac{d^2X}{d\xi^2}-\frac{|\alpha|}{8g^2_{_N}} \xi^2 X = -\frac{C|\alpha|}{g_{_N}}~X~~, 
\end{equation} 
\begin{equation} 
\frac{d^2Y}{d\eta^2}-\frac{|\alpha|}{8g^2_{_N}} \eta^2 Y = -\frac{C|\alpha|}{g_{_N}}~Y~~. 
\end{equation} 
We define 
\begin{equation} 
\gamma^2 =\frac{|\alpha|}{8g_{_N}^2}~~, 
\end{equation} 
\begin{equation} 
E= -\frac{C|\alpha|}{2 g_{_N}}~~, 
\end{equation} 
and thus they retain the same forms as in eqs. (120) and (121) respectively. Focussing on eq.(136) we write it as 
\begin{equation} 
\frac{d^2X}{d\xi^2} -\gamma^2 \xi^2~X = 2~E~X~~. 
\end{equation} 
In terms of $y=\sqrt{2\gamma}~\xi$, eq.(140) becomes 
\begin{equation} 
\frac{d^2X}{dy^2} +(\sigma +\frac{1}{2} -\frac{1}{4}y^2)X =0~~, 
\end{equation} 
where 
\begin{equation} 
\sigma + \frac{1}{2} = -\frac{E}{\gamma}~~. 
\end{equation} 
Eq.(141) has the form of Weber's equation [26] and possesses the following solution 
\begin{equation} 
X_1(y)=D_{_\sigma}(y) = 2^{\frac{\sigma}{2}+\frac{1}{4}}~ y^{-\frac{1}{2}~}W_{\frac{\sigma}{2}+\frac{1}{4}, -\frac{1}{4}}~(y^2/2)~~. 
\end{equation} 
In the above $W_{\mu,\nu}$ is the Whittaker function. Expressing $W_{\mu,\nu}$ in terms of the confluent hypergeometric function we can write 
 
\begin{eqnarray} 
X_1(y) 
&=& \frac{\Gamma(\frac{1}{2})~2^{\frac{\sigma}{2}}}{\Gamma(\frac{1}{2}-\frac{\sigma}{2})}~e^{-\frac{y^2}{4}}~F 
\Big(-\frac{\sigma}{2},\frac{1}{2},\frac{y^2}{2}\Big) \nonumber \\ 
& + &  \frac{\Gamma(-\frac{1}{2})~2^{\frac{\sigma}{2}-\frac{1}{2}}}{\Gamma(-\frac{\sigma}{2})}~y~~e^{-\frac{y^2}{4}}~F 
\Big(\frac{1-\sigma}{2},\frac{3}{2},\frac{y^2}{2}\Big)~~. 
\end{eqnarray} 
For the second solution of eq.(141) we note that from eq.(143) giving the 
relationship betwen $D_\sigma$ and the Whittaker function, we know that 
$D_{-\sigma-1}(\pm iy)$ are solutions linearly independent of $D_{\sigma}(y)$ 
  as $W_{-\frac{\sigma}{2}-\frac{1}{4}, -\frac{1}{4}}(-\frac{y^2}{2})$ is 
  linearly independent of $W_{\frac{\sigma}{2}+\frac{1}{4}, 
    -\frac{1}{4}}(\frac{y^2}{2})$. 
From the asymptotic behaviour of the confluent hypergeometric function we 
deduce that as $y\rightarrow\infty$. 
\begin{equation} 
X_1(y) \sim e^{-\frac{y^2}{4}} y^{\sigma}~~. 
\end{equation} 
i.e., $X_1\rightarrow 0$. For $y\rightarrow -\infty$ we have 
\begin{equation} 
X_1(y) \sim -\frac{(2\pi)^{1/2}}{\Gamma(-\sigma)} e^{\sigma\pi i} 
e^{\frac{y^2}{4}}~y^{-\sigma-1}~~. 
\end{equation} 
i.e., $X_1 \rightarrow \infty$ unless $\sigma$ is a positive integer or zero 
in which case the R.H.S of eq.(146) vanishes. In fact we have the relationship 
\begin{equation} 
D_n(y) = 
2^{-\frac{n}{2}}~e^{-\frac{y^2}{4}}~H_n\Big(\frac{y}{\sqrt{2}}\Big)~~,~~~~~~~\text{n=0,1,2,.....} 
 \end{equation} 
that expresses the parabolic cylinder functions $D_n$ in terms of the Hermite 
polynomials $H_n$. Going back to eqs.(138), (139) and (142) with $\sigma=n$ we 
obtain 
\begin{equation} 
C_n=\frac{n+\frac{1}{2}}{\sqrt{2}|\alpha|^{\frac{1}{2}}}~~, 
\end{equation} 
as the value of the separation constant. The functions $X_n{(\xi)}$ that solve 
eq.(140) are then precisely those that describe the one-dimensional quantum 
oscillator. We write 
 
\begin{equation} 
X_n{(\xi)}=\Big(\frac{\gamma}{\pi}\Big)^{\frac{1}{4}}~\frac{1}{\sqrt{2^n~n!}}~H_n(\sqrt{\gamma}~\xi)~e^{-\frac{1}{2}\gamma\xi^2}~~. 
\end{equation} 
The solutions $Y_n(\eta)$ are identical in form and we obtain for the 
normalized wave function the following 
\begin{equation} 
\psi_n(\xi,\eta)=\Big(\frac{\gamma}{\pi}\Big)^{\frac{1}{2}}~\frac{1}{2^n~n!} ~ 
H_n(\sqrt{\gamma}\xi)~H_n(\sqrt{\gamma}~\eta)~e^{-\frac{1}{2}\gamma(\xi^2+\eta^2)}~~. 
\end{equation} 
For $\sigma \neq n$ the wavefunctions will not have finite norm and solutions 
of eqs.(136) and (137) of the type given in eq.(144) would have to be 
superposed, just as wave packets are constructed in quantum mechanics, in 
order to obtain wave functions capable of describing physical states.

\newpage

\section{Conclusions} 
 
In this work we studied $f(R)$ theories of gravity in two-dimensional 
spacetime with focus on applications to cosmology.  With the metric taken to 
have to the FRW form we were able to obtain solutions for the cosmological 
field equations in the case of pure matter or radiation dominated universe 
when $f(R) = R + \alpha R^n$.  The remarkable feature of these solutions is 
that they readily describe an accelerating universe in contrast to the 
standard FRW cosmology of four-dimensional general relativity.  The horizon 
problem is also readily solved.  As we have stated in sect. 2, the time 
evolution of the scale factor is not affected by the value of the curvature 
constant $k$.  We have also seen that the solution for the radiation dominated 
universe, and one solution for the case of pure matter domination, describe a 
hot big bang.  However an interesting solution in a matter dominated universe, 
given in eq.\eqref{eq:44}, describes a universe that kicks off with an 
infinite size and zero temperature at the start of matter dominance.  It 
subsequently collapses to a finite size and then begins to expand.\\  
Now as we mentioned before, the interest behind the pursuit of $f(R)$ theories 
is partially due to the desire to obtain a description of inflation without 
the introduction of scalar fields.  This is done by seeking solutions to the 
cosmological field equations with the energy-momentum tensor set equal to 
zero, [5].  In sect. 5 we obtained such solutions that characterize power law 
inflation.  Furthermore, with inflation presumed to last for a short period of 
time, we obtained for the case $n=2$ a solution for $t$ near $t_i$, the 
instant of onset of inflation.  This solution displayed exponential dependence 
on time.  For this case we computed the duration of inflation and the number 
of $e$-foldings as well as an estimate for the change that ensues in the value 
of the Hubble parameter from the start to the end of inflation.  The basic 
distinguishing feature between power law and exponential inflation appears to 
be in the behavior of the Ricci scalar.  Exponential inflation obtains when we 
assumed that $R(t)$ can be expanded in a power series about $t=t_i$  with 
finite coefficients.  In particular $R(t_i)$  and  $\dot R(t_i)$ are finite. 
On the other hand in the case of power law inflation these quantities exhibit 
singular behavior at $t=\bar t$.  Another characteristic of our inflationary 
solutions is that they do not depend on the parameter `1 that appears in 
eq.\eqref{eq:31} for $f(R)$.  This is in contrast to the inflationary 
solution in four-dimensional $f(R)$ theories where $n = 2$ describes the 
Starobinsky model [18].  In that case with $\alpha$ written as 
$\alpha=\frac{1}{6M^2}$, where the constant $M $ has the dimension of mass, 
exponential inflation is obtained with $a$, $H$ and $R$ all depending 
on $M$, [5].\\ 
Interest in two-dimensional theories stems partially from the desire to 
investigate the quantum theory in a simple setting. Hence we carried out 
quantization of the theroy in the case of $n=2$. The Wheeler-DeWitt equation 
was derived and its solutions obtained. We were able to solve the equation 
exactly in the entire domain of the variables, unlike the situation in the 
four-dimensional case [19],[27]. Interestingly we found that for $\alpha>0$ 
the equation for the wave function coincided with that of the inverted 
oscillator. For $\alpha<0$ the wave function, under certain conditions, turned 
out to be a product of two quantum harmonic oscillator wave functions in the 
variables $\xi = R+a$ and $\eta=R-a$. 
In conclusion we have studied some aspects of classical and quantum cosmology in two-dimensional 
$f(R)$  theories.  Clearly a lot more issues need to be investigated and we 
hope to return to them in the near future.  
\newpage 
\hglue-0.57cm{\bf REFERENCES}\\ 
 
\hglue-0.57cm$[1]$ H. Weyl, {\it Ann. Phys}; {\bf 59} (1919) 101\\ 
 
\hglue-0.57cm $[2]$  A.S. Eddington, {\it The Mathematical Theory of Relativity} (Cambridge 
University Press, Cambridge) 1923.\\ 
 
\hglue-0.57cm $[3]$  R. Utiyama,  and B.S. Dd Witt,  {\it J. Math. Phys.,} {\bf 3} (1962) 
608; K.S. Stelle, {\it Phys. Rev. D.} {\bf 16} (1977) 953.\\ 
 
\hglue-0.57cm $[4]$  N.D. Birrell, and  P.C.W. Davies , {\it Quantum Fields in Curved 
  Spacetime} (Cambridge University Press, Cambridge) 1982.  I.L. Buchbinder, 
  S.D. Odinstov and I.L. Smapiro,  {\it  Effective Actions in Quantum Gravity} 
  (IOP Publishing, Bristol) 1992. G.A. Vilkovisky, {\it Class. Quant. Grav.,} 
  {\bf 9}(1992) 895.  \\ 
 
\hglue-0.57cm $[5]$ For a review and an exhaustive list of references see.  A. De FELICE and S. Tsujikawa, {\it Living Rev. Relativity,} {\bf 13} (2010), 3.  \\  
Also see: T.P. Sotiriou, arXiv: 0805.1726 preprint (2010). \\ 
 
\hglue-0.57cm $[6]$  L.P. Eisenhart, {\it Riemannian Geometry} (Princeton University Press, 
Princeton) 1949.\\ 
 
\hglue-0.57cm $[7]$  For a review see,  for example, D. Grumiller,  W. Kummer and 
D.V. Vassilevich, {\it Phys. Rep.,} {\bf  36} (2002) 104012.\\
 
\hglue-0.57cm $[8]$   C.G. Callan, S.B. Giddings, J.A. Harvey and A. Strominger,  {\it
  Phys. Rev D,} {\bf  45} (1995) R 100J.\\
 
\hglue-0.57cm $[9]$  G. Mandal,  A.M. Sengupta and S.R. Wadia, {\it Mod. Phys. Lett. A,}
{\bf  6} (1991) 1685.\\
 
\hglue-0.57cm $[10]$  E. Witten, {\it  Phys. Rev. D,} {\bf 44}  (1991) 314. \\
 
\hglue-0.57cm $[11]$  M.A. Ahmed, {\it Phys. Rev. D,} {\bf  61}  (2000) 104012.\\
 
\hglue-0.57cm $[12]$  M.A. Ahmed,  {\it Il Nuovo Cimento,} {\bf  121} (2006) 661.\\
 
\hglue-0.57cm $[13]$  WE follow the conventions of S. Weingberg, {\it Gravitation and
  Cosmology: Principles and Applications of the General Theory of Relativity}
(John Wiley, New York) 1972.\\
 
\hglue-0.57cm $[14]$  C.K. Chan and R.B. Mann, {\it Class. Quant. Grav.,}  {\bf 10} (1993)
913.\\
 
\hglue-0.57cm $[15]$  A. Vilenkin, {\it Phys. Rev. D,} {\bf 32} (1985) 2511.\\
 
\hglue-0.57cm$[16]$  M.P. Hobson, G. Efstathiou  and A.N. Lasenby, {\it  General
  Relativity: An Introduction for Physicists} (Cambridge University Press,
Cambridge) 2006.\\
 
\hglue-0.57cm $[17]$  A.R. Liddle and D.H. Lyth, {\it Cosmological Inflation $\&$ Large
  Scale Structure} (Cambridge University Press, Cambridge) 2000.\\
 
\hglue-0.57cm $[18]$  A.A. Starobinsky, {\it Phys. Lett. B, } {\bf 91} (1980) 99.\\
 
\hglue-0.57cm $[19]$  S.W.Hawking and J.C. Lutrell, {\it Nuclear Phys,}{\bf B247} (1984) 25.\\
 
\hglue-0.57cm $[20]$  W.W. Ford, D.L. Hill, M. Wakano and J.A. Wheeler,{\it Ann. Phys,}{\bf
  7} (1959) 239.\\
 
\hglue-0.57cm $[21]$  W.A. Friedman and C.J. Goebel, {\it Ann. Phys,}{\bf 104} (1977) 145.\\
 
\hglue-0.57cm $[22]$  G.Barton, {\it Ann. Phys,}{\it 166} (1986) 322.\\
 
\hglue-0.57cm $[23]$  N.L. Balazs and A. Voros, {\it Ann. Phys,}{\bf 199} (1990) 123.\\
 
\hglue-0.57cm $[24]$  D. Chruscinski,{\it J.Math Phys,}{\bf 45} (2004) 841.\\
 
\hglue-0.57cm $[25]$  M. Abramowitz and I. Stegun, {\it Handbook of mathematical Functions}
( Dover, New York) 1972.\\
 
\hglue-0.57cm $[26]$  Z.X. Wong and D.R. Guo, {\it Special Functions}(World
Scientific, Singapore) 1989.\\
 
\hglue-0.57cm $[27]$  A.Vilenkin, {\it Phys.Rev,}{\bf D32} (1985) 2511.

\end{document}